\documentstyle[epsf,prl,aps]{revtex}

\begin{document}
\draft
\tighten
\twocolumn[\hsize\textwidth\columnwidth\hsize\csname @twocolumnfalse\endcsname
\title{Optimal Fluctuations and Tail States of non-Hermitian Operators}
\author{A.~V.~Izyumov and B.~D.~Simons}
\address{Cavendish Laboratory, Madingley Road, Cambridge, CB3 0HE, UK}
\date{April 15, 1999}
\maketitle

\begin{abstract}
We develop a general variational approach to study the statistical properties 
of the tail states of a wide class of non-Hermitian operators. The utility of 
the method, which is a refinement of the instanton approach introduced by 
Zittartz and Langer, is illustrated in detail by reference to the problem of 
a quantum particle propagating in an imaginary scalar potential.
\end{abstract} 

\vskip2pc] \narrowtext

Over recent years considerable interest has been shown in the spectral
properties of random Fokker-Planck operators, and their application to the 
dynamics of various classical systems~\cite{isichenko,bouchaud_georges}. Of 
these, perhaps the best known example is the ``Passive Scalar'' problem which 
concerns the diffusion of a classical particle subjected to a random velocity 
field. Here, in contrast to quantum mechanical evolution, the random 
classical dynamics is typically specified by a linear {\em non-Hermitian} 
operator. Non-Hermitian operators also appear in a number of problems in 
statistical physics. For example, the statistical mechanics of a repulsive 
polymer chain can be described in terms of the classical diffusion of 
a particle subject to a random imaginary scalar 
potential~\cite{edwards,kleinert}. Similarly, the statistical mechanics of 
flux lines in a type II superconductor pinned by a background of impurities can
be described as the quantum evolution of a particle in a disordered 
environment subject to an imaginary vector potential~\cite{hatano_nelson}. 
While these connections have been known for a long time, the ramifications of 
non-Hermiticity on the nature of the dynamics is only now being fully explored.
Beginning with early work on random matrix ensembles~\cite{Ginibre,Girko},
a number of attempts have been made to analyze spectral properties of 
non-Hermitian operators, and to apply the results to the description of 
classical systems~\cite{hatano_nelson,sommers88,Haake92,Feinberg97,efetov,%
fyodorov97,chalker_wang,janik,brouwer,mudry_simons,chalker_mehlig,%
izyumov_simons,yurkevich}. For 
example, it has been found that the generic localization properties of 
non-Hermitian systems are drastically different from those of their Hermitian 
counterparts, a fact that can be attributed to an implicit chiral symmetry of 
non-Hermitian operators~\cite{Miller,hatano_nelson,mudry_simons}.

Previous studies of non-Hermitian operators have been largely based on 
perturbative schemes, such as the self-consistent Born approximation in the 
diagrammatic analysis (e.g. Ref.~\cite{chalker_wang}), or the mean field 
approximation in the field-theoretic approach (e.g. 
Ref.~\cite{izyumov_simons}). However, there are 
indications~\cite{balagurov_vaks,samokhin,shnerb} that an important role can 
be played by those parts of the spectrum which are populated by exponentially 
rare, localized states. These ``Lifshitz tail'' states, first introduced 
in the context of semiconductor physics, can not be treated 
perturbatively. Since the original works of Lifshitz~\cite{lifshitz}, a 
number of sophisticated mathematical methods to deal with tail states have 
appeared \cite{halperin_lax,gredeskul}, 
including the instanton technique in statistical field 
theory~\cite{zinn-justin}. The aim of this letter is to propose a 
new non-perturbative scheme to investigate properties of the 
{\em tail states of non-Hermitian operators}. 

Although our approach is quite general, for clarity, we 
choose to explain the main features of the method by applying it to possibly 
the simplest model system. Specifically, we study the Hamiltonian of a 
quantum particle subject to a random imaginary scalar potential
\begin{equation} \label{1.1}
\hat H = -\Delta + i \, V ({\bf r}),
\end{equation}
where the potential $V({\bf r})$ is drawn from a Gaussian $\delta$-correlated 
impurity distribution with zero average, and correlator given by 
$\langle V ({\bf r}) V({\bf r}') \rangle = \gamma \delta ({\bf r} - 
{\bf r}')$. Previous studies have shown that, when averaged over realizations 
of $V({\bf r})$, the Feynman propagator of the Hamiltonian above (\ref{1.1}) 
can be identified with the partition function of a self-repelling polymer 
chain with a contact interaction~\cite{edwards,kleinert}. Moreover, this 
model can be used to describe NMR in inhomogeneous materials, where the 
interplay of diffusion and local variations in the spin precession rate 
results in the dynamics of magnetization being specified by the operator 
(\ref{1.1})~\cite{kiselev_posse}. Indeed, in the case of the latter, the 
tails of the operator (\ref{1.1}) can be shown to provide a significant 
contribution to the relaxation of the NMR signal. 
\begin{figure}[b]
\centerline{\epsfxsize=2.9in\epsfbox{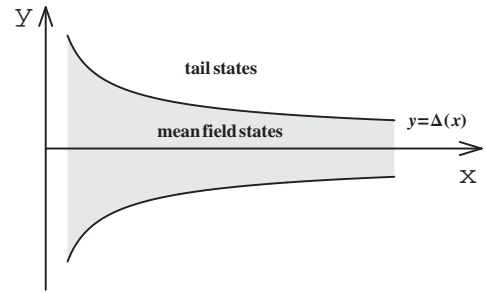}}
\smallskip
\caption{Support of the states in the complex plane ($d=1$). The density of 
states outside the shaded region is zero if calculated in the self-consistent 
Born approximation.}
\end{figure}
The Hamiltonian (\ref{1.1}) is non-Hermitian, and its eigenvalues $\epsilon_k$ 
occupy some area in the complex plane. In the self-consistent Born approximation,
the density of complex eigenvalues, defined as 
$\rho =\langle \sum_k\delta (x-x_k) \delta (y-y_k)\rangle$, 
is given by~\cite{izyumov_simons}
\begin{equation} \label{1.5}
\rho(x,y)=\cases{(4\pi\gamma)^{-1}, & $|y|<\Delta (x)$, \cr 0 , & $|y|>\Delta
(x)$,}
\end{equation}
\noindent
where $x$ and $y$ are the real and imaginary parts of the complex energy
$\epsilon = x+iy$, $\Delta (x) = 2\pi\gamma \nu (x)$, and $\nu (x)$ 
is the density of states of the clean ($\gamma = 0$) system. 
The validity of this ``mean field'' result is restricted to values of $x$ such
that $\Delta(x)\ll x$, which corresponds to large $x$ in dimensions lower than
four.

According to Eq.~(\ref{1.5}), there are no states outside the region
$|y| < \Delta(x)$. This can be understood if one recalls that the mean field 
approximation is essentially a way to determine the self-consistent 
contribution of typical fluctuations of the random potential.
On the other hand, numerical simulations \cite{izyumov_simons} suggest the
existence of rare states with $|y| \gg \Delta(x)$, which are generated
by atypically strong fluctuations of $V({\bf r})$. Such fluctuations occur 
with an exponentially small probability $\exp[-W/\gamma]$, where 
\begin{equation} \label{2.4}
W=\frac{1}{2}\int V^2({\bf r}) d{\bf r}.
\end{equation}
For any given configuration $V({\bf r})$ of the random potential, we will 
describe $W$ as the {\em energy} associated with that configuration. This 
should not be confused with the ``energy'' $\epsilon$ of a quantum state, 
which by definition is the eigenvalue of $\hat H$ corresponding to that 
state. Since strong fluctuations of the disorder potential are suppressed 
by an exponential factor, states with $|y| \gg \Delta(x)$ are dominated by 
those configurations of $V({\bf r})$ that have the highest statistical weight 
$\exp[-W/\gamma]$ or, equivalently, the lowest energy $W$. Thus, we come to 
the idea of the {\em optimal fluctuation method}, whose rigorous formulation 
can be given as follows: Amongst all the configurations of $V({\bf r})$ such 
that $\epsilon$ is an eigenvalue of the Hamiltonian (\ref{1.1}), we choose 
the one that minimizes the energy functional 
$W[V({\bf r})]$. The density of states is then given with exponential accuracy 
by the statistical weight of the minimal configuration, $\rho(\epsilon) \sim
\exp[-W_{min}(\epsilon)/\gamma]$. This approach is similar to that employed 
in the seminal work by Zittartz and Langer~\cite{zittartz_langer} in the 
treatment of the Hermitian model. There, a saddle-point technique was applied 
to estimate the functional integral over 
$V({\bf r})$. Although based on the same physical ideas, this technique leads 
to inconsistent results when formally applied to a non-Hermitian problem. 
Instead, we introduce below a more general variational formulation tailored 
to the consideration of non-Hermitian operators. 

The arguments outlined above are not restricted to the particular form of the 
Hamiltonian (\ref{1.1}), and can be applied whenever one has to deal with 
tail states (i.e. {\em localized} states created by strong fluctuations of 
the disorder potential). Formally, the limitations of this scheme are set by 
the inequality $W_{min}(x,y) \gg \gamma$, which defines a certain area in 
the complex plane of eigenvalues. Below we will show that the precise form 
of the optimal fluctuation potential can, in general, be determined by 
solving a system of coupled non-linear equations. For the particular case of
the Hamiltonian (\ref{1.1}), we will obtain an explicit form of the solution
in the limit $y \ll x$.

With this preparation, we now turn to the derivation of the variational 
approach: Our aim is to minimize the functional $W[V({\bf r})]$ subject to
the constraint $\mbox{det} (\hat H - \epsilon) = 0$. Account for the latter
can be made by introducing two Lagrange multipliers, $\mu_1$ and $\mu_2$,
which in turn leads to the functional 
\begin{equation} \label{3.2}
F[V({\bf r})] = W - \mu_1 \, \mbox{Re} \, \mbox{det}(\hat H -
\epsilon) - \mu_2 \, \mbox{Im} \, \mbox{det}(\hat H - \epsilon).
\end{equation}
If $\epsilon$ is an eigenvalue of $\hat H$ with left and right eigenfunctions
$\psi^{L}({\bf r})$ and $\psi^{R}({\bf r})$, a spectral decomposition of the 
Hamiltonian (\ref{1.1}) obtains the identity
\begin{equation} \label{3.7}
\frac{\delta}{\delta V({\bf r})} \mbox{det}(\hat H - \epsilon) = i \,
\psi^{L*}({\bf r}) \psi^{R}({\bf r}) \prod_{k, \epsilon_k \not = \epsilon}
(\epsilon_k - \epsilon),
\end{equation}
where $\epsilon_k$ denote the remaining eigenvalues of $\hat H$ (i.e. those 
different from $\epsilon$). Equating the functional derivative $\delta F / 
\delta V({\bf r})$ to zero, and applying Eq.~(\ref{3.7}) we obtain
\begin{equation} \label{3.10}
V({\bf r}) =  \lambda_1 \, \mbox{Re} \, \psi^{L*}({\bf r}) \psi^{R}({\bf r})
+ \lambda_2 \, \mbox{Im} \, \psi^{L*}({\bf r}) \psi^{R}({\bf r}),
\end{equation}
where $\lambda_1$ and $\lambda_2$ denote redefined Lagrangian multipliers.
When combined with the eigenvalue equations
\begin{equation} \label{3.8}
\hat H \psi^{R}({\bf r}) = \epsilon \, \psi^{R}({\bf r}), \quad
\hat H^{\dag} \psi^{L}({\bf r}) = \epsilon^* \, \psi^{L}({\bf r}),
\end{equation}
and with the binormality condition
\begin{equation} \label{3.9}
\int d {\bf r} \, \psi^{L*}({\bf r}) \psi^{R}({\bf r}) = 1,
\end{equation}
Eq.~(\ref{3.10}) defines a closed system of non-linear equations for 
$\psi^R$, $\psi^L$, $\lambda_1$, $\lambda_2$, which has to be solved in order 
to extract the optimal configuration $V({\bf r})$ of the disorder potential.

Note that Eq.~(\ref{3.10}) depends explicitly on the nature of the impurity
distribution and on the structure of the operator at hand, whereas 
Eqs.~(\ref{3.8}) and (\ref{3.9}) are universal. To treat other random 
non-Hermitian operators, one simply has to modify Eq.~(\ref{3.10}) 
accordingly. For example, in the case of the Passive Scalar problem, where 
the disorder has the form of a random velocity field, the analogue of 
Eq.~(\ref{3.10}) relates the optimal configuration of the random velocity field
to spatial derivatives of the eigenfunctions $\psi^R$, 
$\psi^L$~\cite{izyumov_ruban}. 

For the Hamiltonian (\ref{1.1}), further progress can be made by exploiting 
symmetry properties. From the relation $\hat H^{\dag}=\hat H^*$ it follows 
that one can constrain the complex wavefunctions to obey the relation 
$\psi^{R}({\bf r})=\psi^{L*}({\bf r})=\psi ({\bf r})$. In doing so, leaving
aside the lengthy but straightforward analysis~\cite{izyumov_ruban}, one can 
show that the following simplifications
obtain: Firstly, Eq.~(\ref{3.9}) can be effectively 
disregarded; secondly, by performing a complex rotation of $\psi({\bf r})$, 
one finds that arbitrary non-zero values can be assigned to the Lagrange 
multipliers $\lambda_1$ and $\lambda_2$. For convenience, we will set 
$\lambda_1=1$, $\lambda_2=0$. With these simplifications, focusing initially
on the one-dimensional problem, one finds
\begin{equation} \label{4.2}
V = \mbox{Re}\, \psi^2,
\end{equation}
where the wavefunction $\psi$ is obtained self-consistently by solving
the Schr\"odinger equation in the potential $V$,
\begin{equation} \label{4.1}
-\psi'' + i V \psi = \epsilon \psi.
\end{equation}
Eqs.~(\ref{4.2}) and (\ref{4.1}), which represent the non-Hermitian analogue 
of the non-linear Schr\"odinger equation of Zittartz and Langer, are the main 
result of the paper. To complete the program, one should find the localized 
solution $\psi$ (which is assumed to be unique) of Eqs.~(\ref{4.2}) and 
(\ref{4.1}) for each value of $\epsilon$, and calculate the energy 
$W(\epsilon)$ of the corresponding configuration $V = \mbox{Re}\, \psi^2$ of 
the potential. Note that, in the derivation of Eqs.~(\ref{4.2}) and 
(\ref{4.1}), no approximation has been made. However, the applicability of 
the optimal fluctuation method {\em itself} is restricted by the condition 
$W_{min}(\epsilon) \gg \gamma$. 

As mentioned above, an analogous calculation can be carried out for other 
model systems including the Passive Scalar operator
\begin{equation} \label{fokker}
\hat {\cal H}_{\rm ps}=-\Delta + {\bf v}({\bf r})\cdot \nabla.
\end{equation}
Indeed, in the case of a Gaussian distributed incompressible ($\nabla \cdot 
{\bf v} = 0$) flow, the tail states of $\hat {\cal H}_{\rm ps}$ are found to
be governed by equations which have the {\em same} form as Eqs.~(\ref{4.2}) 
and (\ref{4.1})~\cite{izyumov_ruban}. 

In the analysis of Eqs.~(\ref{4.2}) and (\ref{4.1}), it is convenient to 
interpret the one-dimensional spatial coordinate $r$ as a time
$t$, and the complex wavefunction $\psi(t)$ as the position of a fictitious 
classical particle in the two-dimensional plane. With this interpretation, 
one can recast Eqs.~(\ref{4.2}) and (\ref{4.1}) in the Lagrangian form,
$(\partial{\cal L}/\partial \psi'_{1,2})'=\partial{\cal L}/\partial\psi_{1,2}$,
where
\begin{equation} \label{4.6}
{\cal L}=\psi'_1\psi'_2-x\psi_1\psi_2-\frac{y}{2}(\psi^2_1- \psi^2_2)+
\frac{1}{4}(\psi^2_1-\psi^2_2)^2,
\end{equation}
and $\psi = \psi_1 + i \psi_2$ has been separated into real and imaginary 
parts. The Lagrangian (\ref{4.6}) has at least one invariant of the motion 
--- the classical energy. Although we can not rule out the possibility of this
system being integrable, we have been unable to find a second invariant
of the motion. Therefore, integrability remains an open question, which 
deserves a separate investigation.

As follows from Eqs.~(\ref{4.2}) and (\ref{4.1}), the wavefunction of a tail 
state is defined by the components of the energy, $x$ and $y$, which 
determine the position of the state in the complex plane of eigenvalues. In 
order to exploit the optimal fluctuation method to its fullest potential, one 
should analyze these equations over the entire complex plane. However, at 
present, we do not have a method of finding the exact analytical solution of 
Eqs.~(\ref{4.2}) and (\ref{4.1}) for arbitrary values of $x$ and $y$. 
Therefore, we will focus on a specific domain, in which an approximate 
solution can be found. 

As we will see later, the density of Lifshitz tails decays exponentially as a
function of the distance from the boundary of the ``mean field spectrum'',
$|y|=\Delta(x)$. Therefore, the most physically relevant states lie close to 
the boundary. For such states, the condition $|y| \ll x$ is satisfied 
(recall that $\Delta(x) \ll x$) allowing the following separation of scales: 
Since the wavefunction of a tail state oscillates with a high frequency 
$\sqrt{x}$, while the amplitude of these oscillations varies at a much lower 
rate $|y|/\sqrt{x}$ (the ratio of the two time-scales being $|y|/x \ll 1$), 
the solution of Eqs.~(\ref{4.2}) and (\ref{4.1}) can be parameterized in the 
form of a wavepacket,
\begin{equation} \label{wavepacket}
\psi(t)=\sqrt{2|y|} \left [
\varphi_+ \left ( \frac{|y|t}{2\sqrt{x}} \right ) e^{i\sqrt{x}t} +
\varphi_- \left ( \frac{|y|t}{2\sqrt{x}} \right ) e^{-i\sqrt{x}t}
\right ].
\end{equation}

Substituting the parameterization (\ref{wavepacket}) into Eqs.~(\ref{4.2}) 
and (\ref{4.1}), one obtains an equation for the envelope. Following a 
standard procedure~\cite{zakharov}, we neglect the second derivatives of 
$\varphi_{\pm}$, and perform the averaging over the intermediate scales. As a 
result, one obtains a system of {\em parameter-free} first-order differential 
equations for $\varphi_{\pm}(\tau)$:
\begin{equation} \label{5.4}
\begin{array}{rccl}
\varphi'_+&=&-&\varphi_+ + 5 \varphi_+^2\varphi_- + \varphi_-^3,\\
\varphi'_-&=& &\varphi_- - 5 \varphi_+\varphi_-^2 - \varphi_+^3.
\end{array}
\end{equation}
In general, the solution of equations of this kind, which can easily be
computed numerically, yields a universal dimensionless constant in the final
result for the density of states. Surprisingly, for the problem at hand,
Eqs.~(\ref{5.4}) can be represented in the form of Hamilton's equations,
wherein $\varphi_{\pm}$ play the role of the canonical variables,
$\varphi'_\pm = \pm \partial {\cal H} / \partial \varphi_\mp$,
%
%
with the Hamiltonian
\begin{equation} \label{6.2}
{\cal H}=-\varphi_+\varphi_- +\frac{5}{2}\varphi^2_+\varphi^2_-
+\frac{1}{4}(\varphi^4_+ + \varphi^4_-).
\end{equation}
Of the infinite set of solutions, defined by values of ${\cal H}$, we are
interested in the localized solution, which corresponds to
${\cal H} = 0$. Integrating Eqs.~(\ref{5.4}) along the line ${\cal H} = 0$,
we obtain
\begin{equation} \label{6.6}
\varphi_\pm (\tau) = \frac{2 e^{\pm \tau}}{\sqrt{e^{4\tau}+e^{-4\tau}+10}}.
\end{equation}
Now, with the help of Eq.~(\ref{wavepacket}), we can restore the wavefunction
$\psi(t)$. It has a rather unusual shape shown in Fig.~2(a). The combination
of an oscillatory wavefunction modulated by a localized envelope is 
characteristic of the tail states of non-Hermitian operators. The optimal 
configuration of $V(t)$,  which is expressed through $\psi(t)$ according to 
Eq.~(\ref{4.2}), also has the form of localized oscillations (see Fig.~2(b)), 
and its energy can be straightforwardly determined:
\begin{equation} 
W_{min}(\epsilon) = \alpha |y| \sqrt{x}, 
\quad 
\alpha = \frac{4}{\sqrt{6}} \log (5 + 2 \sqrt{6}) \approx 3.744.
\end{equation}
Recalling that, in one dimension, $\nu(x)=1/(2\pi\sqrt{x})$,
we  arrive at the expression for the density of tail states,
\begin{equation} 
\rho(\epsilon) \sim \exp \left [ -\alpha \frac{|y|}{\Delta(x)} \right ],
\quad \Delta (x) \ll |y| \ll x.
\end{equation}

We remark that, in higher dimensions, the spherical symmetry of the optimal 
potential reduces the problem to being effectively one-dimensional. In this 
case, one obtains the general result~\cite{izyumov_ruban},
\begin{equation}
\rho(\epsilon) \sim 
\cases{
\exp\left [-\alpha\left (\frac{x}{|y|}\right )^{d-1}
\frac{|y|}{\Delta(x)} \right ]  , & $\Delta (x) \ll |y| \ll x$, \cr 
\exp\left [-\beta\frac{|y|}{x}
\frac{|y|}{\Delta(x)} \right ]  , & $|y| \gg  x$,\cr}
\end{equation}
where the values of the numerical factors $\alpha$ and $\beta$ depend on the 
dimensionality, and $\Delta (x) = 2\pi\gamma\nu (x) \sim \gamma x^{(d-2)/2}$.
\begin{figure}[t]
\centerline{\epsfxsize=2.6in\epsfbox{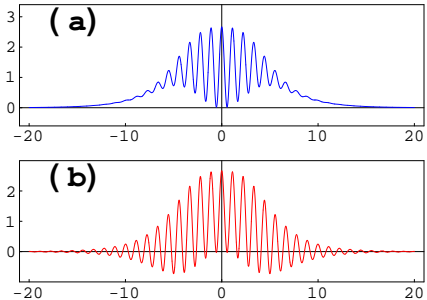}}
\smallskip
\caption{Solution of the non-linear Scr\"odinger equation (\ref{4.2}) and 
(\ref{4.1}) for $x=8.0$, $y=1.0$: (a) shows the square modulus $|\psi|^2$
of the complex wavefunction. 
The optimal fluctuation of the disorder potential $V = \mbox{Re}\, \psi^2$ 
is shown in (b).}
\end{figure}

For completeness, one should add that in addition to the tail states 
depicted in Fig.~1 and studied here, the random imaginary scalar potential 
exhibits another class of exponentially rare states which inhabit the region 
of very small $x \rightarrow 0$~\cite{izyumov_simons}. The origin of the 
latter seems to be associated with large areas in real space which are almost 
free of disorder. Such states are inaccessible within the present framework.

To summarize, in this paper we have developed a general variational approach 
to study spectral properties of non-Hermitian operators. Applied to the 
problem of a particle propagating in a random imaginary scalar potential, 
one finds that properties of the strongly localized (tail) states are 
governed by a system of coupled non-linear dynamical equations. These 
equations show that tail states associated with the non-Hermitian Hamiltonian
exhibit oscillations on scales much shorter than the localization length. 
Employing a procedure of scale separation, both the density of tail states 
as well as the corresponding wavefunctions have been determined.

We would like to thank V.~Ruban, K.~Samokhin, I.~Smolyarenko and A.~Moroz for
helpful discussions.


\end{document}